\documentclass[prl,notitlepage,twocolumn]{revtex4}

\usepackage{amsmath}
\usepackage{amssymb}
\usepackage{bm}
\usepackage{graphicx}
\usepackage{times}

\usepackage{color}
\usepackage{ulem}  

\begin{document}

\title{All-dielectric resonant meta-optics goes active}

\author{Isabelle Staude$^1$}
\author{Thomas Pertsch$^{1,2}$}
\author{Yuri Kivshar$^3$}

\affiliation{$^1$Institute of Applied Physics, Abbe Center of Photonics, Friedrich Schiller University Jena, 07745 Jena, Germany}
\affiliation{$^2$Fraunhofer Institute for Applied Optics and Precision Engineering, 07745 Jena, Germany}
\affiliation{$^3$Nonlinear Physics Centre, Australian National University, Canberra ACT 2601, Australia}

\begin{abstract}
All-dielectric resonant nanophotonics is a rapidly developing research field driven by its exceptional application potential for low-loss nanoscale metadevices. The tight confinement of the local electromagnetic fields and interferences in resonant photonic nanostructures can boost many optical effects, thus offering novel opportunities for the subwavelength control of light-matter-interactions. Active, light-emitting nanoscale structures are of particular interest, as they offer unique opportunities for novel types of light sources and nanolasers. Here, we review the latest advances in this recently emerged and rapidly growing field of {\it active dielectric resonant nanophotonics} enabled by dipolar and multipolar Mie-type resonances. More specifically, we discuss how to employ dielectric nanostructures for the resonant control of emission from quantum dots, two-dimensional transition metal dichalcogenides, and halide perovskites. We also foresee various future research directions and applications of active all-dielectric resonant nanostructures including but not limited to lasing in topologically robust systems, light-matter interactions beyond the electric dipole limit, light-emitting surfaces, and tunable active metadevices.
\end{abstract}

\maketitle

\section{Introduction}

The study of resonant dielectric nanostructures with high refractive index is a new research direction in nanoscale optics and metamaterial-inspired nanophotonics~\cite{kuznetsov,staude,kruk}. The concept of all-dielectric resonant nanophotonics is driven by the idea to employ subwavelength dielectric Mie-resonant nanoparticles as "meta-atoms" for creating highly efficient optical metasurfaces and metadevices. These are defined as devices having unique functionalities derived from structuring of functional matter on the subwavelength scale~\cite{nmat}, often termed as "meta-optics" for emphasizing the importance of optically-induced magnetic response of their artificial subwavelength-patterned structures. Because of the unique optical resonances and their various combinations for interference effects accompanied by strong localization of the electromagnetic fields, high-index nanoscale structures are expected to complement or even replace different plasmonic components in a range of potential applications. Moreover, many concepts which had been developed for plasmonic structures, but fell short of their potential due to strong losses of metals at optical frequencies, can now be realized based on low-loss dielectric structures.

High-index dielectric meta-atoms can support both electric and magnetic Mie-type resonances in the visible and near-infrared spectral range, which can be tailored by the nanoparticle geometry. Mie-resonant silicon nanoparticles have recently received considerable attention for applications in nanophotonics and metamaterials~\cite{staude,kruk} including optical nanoantennas, wavefront-shaping metasurfaces, and nonlinear frequency generation. Importantly, the simultaneous excitation of strong electric and magnetic Mie-type dipole and multipole resonances can result in constructive or destructive interferences with unusual spatial scattering characteristics~\cite{oe_wei}, and it may also lead to the resonant enhancement of magnetic fields in dielectric structures that could bring many novel effects in both linear and nonlinear regimes.

Importantly, tight confinement of the local electromagnetic fields in resonant photonic nanostructures can boost many effects of light-matter interaction, including light emission processes. Such structures can be tuned to optical resonances at the wavelength of the material emission. All-dielectric implementations of light-emitting nanostructures offer several advantages over their plasmonic counterparts, including a high radiation efficiency, versatile options to achieve directional far-field coupling, lower heat generation, and -- depending on the material choice -- CMOS compatibility. Also, the low dissipation losses of higher order resonances mean that detrimental quenching, which is common for plasmonic nanoantennas, is no longer an issue. Furthermore, light sources can be directly placed inside the dielectric material of the nanoresonator, where the near-field enhancement is usually the largest. This also protects the active material from harmful ambient conditions such as chemical pollution or mechanical damages. A common approach is to create active dielectric nanoantennas with optical resonances at the wavelength of the material emission. For example, materials such as GaAs and GaN (and several other III-V semiconductors) possess direct electronic inter-band transitions supporting efficient light emission, while their refractive indices are high enough to exhibit strong Mie resonances in the visible range. Altogether, all-dielectric Mie-resonant nanostructures offer many new opportunities for controlling light emission at the nanoscale, providing a pathway for the realization of novel types of light sources and nanolasers. 
To achieve the desired functionalities, the study of {\it active all-dielectric resonant nanostructures} is of crucial importance. Here we use the term {\it active} in the sense of {\it light-emitting} structures, in close analogy to direct-bandgap (active, light emitting) semiconductors.  

In this Perspective, we review the current progress in this recently emerged field of active dielectric meta-optics driven by multipolar Mie-type resonances and outline possible future developments. We concentrate on light emission originating from electronic transitions within active materials, including spontaneous and stimulated emission processes, while excluding from the discussion the recent research efforts in the area of nonlinear frequency generation in all-dielectric nanostructures. Particularly, we will touch on Mie-resonant nanostructures empowered with quantum dots or nitrogen-vacancy (NV) centers, coupling of Mie resonators to atomically thin transition metal dichalcogenides (TMDCs) or halide perovskites, enhancement of lasing in metasurfaces, as well as tailoring spontaneous emission with Mie-resonant metasurfaces. We envisage various future research directions and applications of this field, including lasing in topologically robust systems, light-matter interactions beyond the electric dipole limit, flat sources of complex light fields, and tunable active metadevices.\\

\section{General Properties}

\subsection{Resonances in dielectric nanostructures} 

Resonances and interference effects play a crucial role in all-dielectric meta-optics because they allow to enhance substantially both electric and magnetic fields and to tailor their coupling with light sources and active centers, being an important requirement for active nanophotonics. Unlike metal-based plasmonic structures, the electromagnetic fields penetrate into dielectric nanostructures. Thus, in order to optimize the field enhancement in dielectric nanostructures, different strategies for the nanostructure design as compared to nanoplasmonics are needed. Popular strategies rely on geometric resonances and interferences between different modes. Based on this idea, we can identify several physical mechanisms for the field enhancement and engineering in all-dielectric meta-optics, such as
\begin{itemize}
\item Mie resonances,
\item Fano resonances,
\item Bound states in the continuum.
\end{itemize}

{\bf Mie resonances} are associated traditionally with the exact Mie solutions of Maxwell's equations and the colors of colloidal solutions of gold nanoparticles. However, optical Mie resonances of high-index dielectric nanoparticles can also be employed for the control of light below the free-space diffraction limit~\cite{kuznetsov,staude,kruk}. Importantly, high-index dielectric nanoparticles of simple geometries can support electric and magnetic type resonances of comparable strengths. A strong {\it magnetic dipole} (MD) resonance occurs due to a coupling of incoming light to the circular displacement currents of the electric field, owing to the field penetration and phase retardation inside the particle. This becomes possible when the wavelength inside the particle becomes comparable to the particle's spatial dimension such as its radius, $2R \approx \lambda/n$, where $n$ is the refractive index of the particle's material, $R$ is the nanoparticle's radius, and $\lambda$ is the wavelength of light inside the material. This is a type of {\it geometric resonance}, which requires that a nanoparticle should have a large refractive index in order to support strong MD resonances for vacuum wavelengths larger than the particle's dimension. Similarly, a dielectric nanoparticle can support electric dipole (ED) and high-order electric and magnetic resonances, i.e. electric quadrupole (EQ), magnetic quadrupole (MQ) etc.. All such resonances can be coupled to external or internal emitters, which can be enhanced further by engineering the interference between different modes. Such Mie resonances are usually associated with large concentration of the incident electromagnetic field inside the particles~\cite{trib_sr} suggesting at least {\it two-orders-of-magnitude enhancement} of many optical effects observed with high-index dielectric nanoparticles, in comparison with non-resonant cases ~\cite{kruk}. 

{\bf Fano resonances} are a fascinating phenomenon of wave physics observed across various branches of optics, including photonics, plasmonics, and metamaterials~\cite{fano_rev}. Nanophotonics is dealing with extremely strong and confined optical fields at subwavelength scales far beyond the diffraction limit. A proper combination of nanoparticles can support Fano resonances arising from interference between different localized modes and radiative electromagnetic waves~\cite{fano_review}. In nanophotonic structures, the Fano resonances typically manifest themselves as resonant suppression of the total scattering cross-section accompanied by an enhanced absorption. The Fano resonances allow to confine light more efficiently, and they are characterized by a steeper dispersion than conventional resonances, which make them promising for nanoscale biochemical sensing, switching or lasing applications.

It was commonly believed that Fano resonances in nanophotonics originate solely from the excitation of plasmonic modes being associated with the electric field resonances and plasmonic “hot spots”. In a sharp contrast to this common belief, recently it was demonstrated experimentally~\cite{small}  that subwavelength nanoparticles with high refractive index can support strong Fano resonances through the interference of optically-induced MD rather than ED resonant modes, as was predicted theoretically~\cite{fano_nl} already before the experimental verification~\cite{small}.

A rigorous analysis~\cite{fano_pra} reveals principal differences between dielectric and plasmonic oligomers, which are symmetric clusters of nanoparticles. First, in comparison with plasmonic oligomers, all-dielectric structures are less sensitive to the separation between the particles, since the field is localized inside the dielectric nanoparticles. Second, dielectric nanostructures exhibit almost negligible losses, which allows observing Fano resonances for new geometries not supported by their plasmonic counterparts. These novel features lead to different coupling mechanisms between nanoparticles making all-dielectric oligomer structures very attractive for various applications in active nanophotonics.

{\bf Bound states in the continuum} originate from strong coupling between the modes in dielectric structures such as photonic crystals, metasurfaces, and isolated resonators~\cite{BIC_review}. They attracted a lot of attention in photonics recently as an alternative mean to achieve very large quality factors ($Q$-factors) for lasing~\cite{BIC_nature} and also tune the system to the regime of {\it a supercavity mode}~\cite{rybin}. 

A true {\it bound state in the continuum} (BIC) is a mathematical object with an infinite value of the $Q$ factor and vanishing resonance width, and it can exist only in ideal lossless infinite structures or for extreme values of parameters~\cite{hsu,alu}. In practice, BIC can be realized as a quasi-BIC mode, also known as a supercavity mode~\cite{rybin}, when both the $Q$ factor and resonance width become finite. Nevertheless, the BIC-inspired localization of light makes it possible to realize high-$Q$ quasi-BIC modes in optical cavities and photonic crystal slabs and coupled optical waveguides.

Importantly, even a single subwavelength high-index dielectric resonator can be tuned into the regime of a supercavity mode. This can be achieved by varying the nanoparticle’s aspect ratio~\cite{prl_kirill}, when the radiative losses are almost suppressed due to the Friedrich-Wintgen scenario of destructive interference of leaky modes~\cite{fri_win}. 

In connection with broader applications, many metasurfaces composed of arrays of dissimilar meta-atoms with a broken in-plane symmetry can support high-$Q$ resonances directly associated with the concept of bound states in the continuum~\cite{new_PRL}. This includes, in particular, broken-symmetry Fano dielectric metasurfaces for enhancement of nonlinear effects~\cite{brener} and recently reported sensing with pixelated dielectric metasurfaces~\cite{science}. All such structures can be employed in active nanophotonics, since they give rise to strong light-matter interaction on the nanoscale.

There exists a direct link between quasi-BIC states and Fano resonances since these two phenomena are supported by the similar physics. Indeed, it has been shown~\cite{new_PRL} that the transmission spectra of dielectric metasurfaces with broken symmetry near the condition of the quasi-BIC resonance can be described explicitely by the classical Fano formula, and the observed peak positions and linewidths correspond exactly to the real and imaginary parts of the eigenmode frequencies. The Fano parameter becomes ill-defined at the BIC condition, which corresponds to a collapse of the Fano resonance~\cite{fonda}.

\subsection{Purcell effect}

\begin{figure*}
\centering
\includegraphics[width=0.9\linewidth] {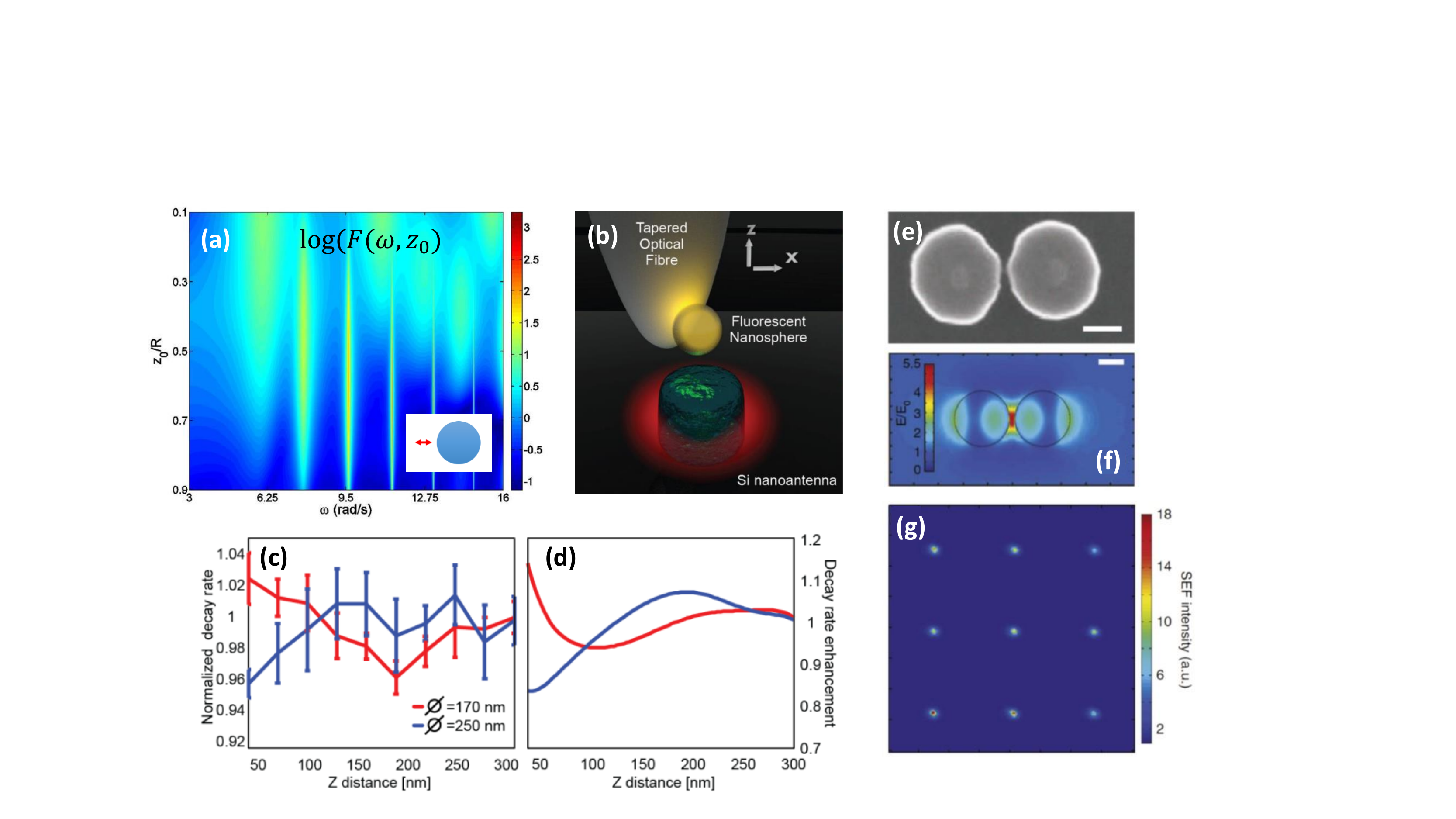}
\caption{{\bf Purcell factor of Mie-resonant dielectric nanoparticles.} (a) Logarithm of the Purcell factor $(log[F(\omega,z_0)])$ as a function of the frequency $\omega$ of the emitter and its position $z_0$ in the z axis for a Si nanosphere having a radius of R=200 nm \cite{Bonod}. The orientation of the dipole is along the z axis. (b) Sketch of the experimental study performed in \cite{Bouchet}: a 100 nm diameter fluorescent sphere grafted at the extremity of a tapered optical fiber, is scanned relative to an Si nanoantenna. Experimental (c) and theoretical (d) evolution of the emission decay rate as a function of the axial distance between the fluorescent nanosphere and the Si nanoantenna illustrated in (b) for two different Si nanoantenna diameters (see inset in (d)). (b-d) taken from \cite{Bouchet}. (e) Top-view scanning-electron micrograph of a single dimer-like Si nanoantenna suggested for hot-spot formation with ultra-low heat conversion \cite{Maier}. Scale bar: 100 nm. (f) Numerically calculated near-field distribution map for the Si dimer nanoantenna excited at resonance, showing the confinement of the electric field in the gap. Maximum enhancement value: 5.5. Scale bar: 100 nm. (g) Experimental map of Nile Red fluorescence enhancement generated by the Si dimer nanoantennas. (e-g) taken from \cite{Maier}}
\label{fig1}
\end{figure*}

One of the important effects in nanophotonics regards the control of electromagnetic emission via the so-called Purcell effect~\cite{Purcell}. During the past decades the research in nanophotonics has been focused on electric dipole emission, however recent progress in the nanofabrication and studies of magnetic quantum emitters have stimulated the investigation of the magnetic side of spontaneous emission~\cite{Krasnok}. Spontaneous emission of both electric and magnetic dipole quantum emitters can be enhanced by the use of various nanophotonics systems, and such an enhancement is associated with both types of electric and magnetic Mie resonances. The Purcell enhancement factor can be calculated theoretically for spherical dielectric resonators by using a modal approach~\cite{Bonod}. The calculated Purcell factor for a spherical Mie resonator (refractive index 3.5, radius 0.2 $\mu$m) embedded in air is presented in Fig.\,\ref{fig1}\,(a). It can be shown that the effective volume associated with each normal mode is related to the translation-addition coefficients of a displaced dipole. For a Si resonator homogeneously doped with electric dipolar emitters, the average electric Purcell factor dominates over the magnetic one.  

\begin{figure*}
\centering
\includegraphics[width=1.0\linewidth] {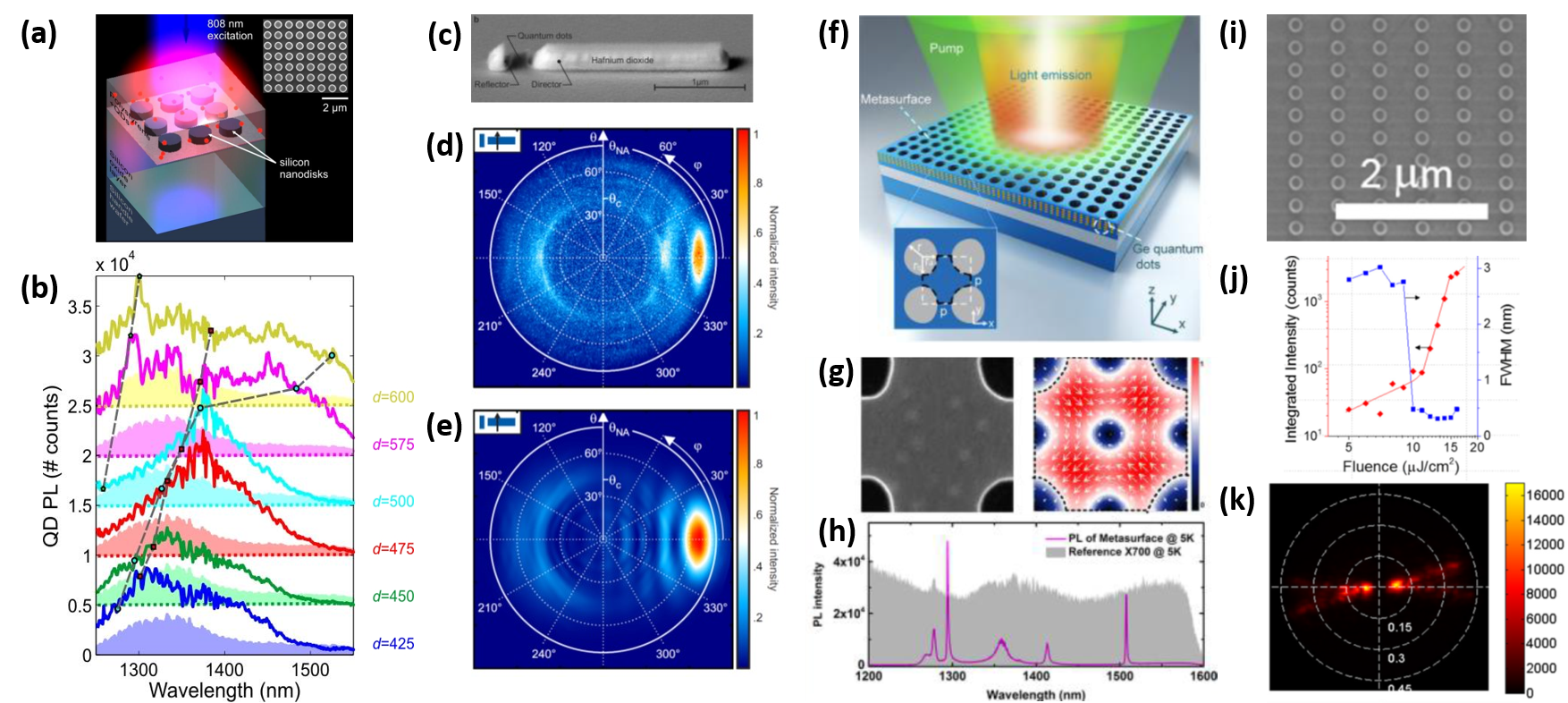}
\caption{{\bf Dielectric nanoantennas and metasurfaces empowered by active semiconductors and quantum dots.} (a) Sketch of a Si nanoantenna array covered by colloidal QDs in a polystyrene matrix. The inset shows a scanning electron micrograph of a typical Si antenna sample. (b) Geometric tuning of the nanoantenna resonance allows for shaping of the emission spectra of the QDs. (a,b) taken from \cite{Staude2015}. (c) Electron microscope image of a hafnium dioxide truncated-waveguide nanoantenna, which was locally functionalized with colloidal QDs. (d) Experimental and (e) numerically calculated back-focal plane image of the emission from the hybrid QD-nanoantenna system, providing clear evidence of directional emission. (c-e) taken from \cite{Peter2017}. (f) Sketch of a Fano-resonant dielectric metasurface consisting of asymmetric holes in a Si slab containing Ge QDs. (g) Close-up electron microscopy image and a calculated near-field profile of a single unit cell of the metasurface. (h) Measured photoluminescence of the metasurface containing Ge QDs, showing three-orders of magnitude enhancement as compared to the unstructured reference wafer. (f-h) taken from \cite{yuan2017}. (i) Top-view electron micrograph of a resonant GaAs nanopillar array employed in lasing experiments. (j) Integrated output intensity of the GaAs nanopillar array and full width at half maximum (FWHM) of the emission peak as a function of input fluence. The observed S-curve in the input-output dependence indicates the transition from spontaneous to stimulated emission. (i-k) taken from \cite{Ha2018}.}
\label{fig2}
\end{figure*}

The emission enhancement due to the Mie resonances has been recorded in a number of experiments on the measurements of the fluorescence rate~\cite{Maier, Bouchet, Aigouy, Zalogina, Regmi} [see Figs.\,\ref{fig1}\,(b-g)].  In particular, Zalogina {\it et al.}~\cite{Zalogina} suggested active nanoantennas based on diamond nanoparticles with embedded nitrogen-vacancy (NV) centers coupled to Mie resonances of nanoparticles, and demonstrated experimentally the enhancement of the fluorescence rate of the emitters due to particle resonances with respect to the non-resonant regime. 

\section{Semiconductors and quantum dots}

Spontaneous or stimulated light emission from semiconductor materials originating from electronic interband transitions via radiative channels are at the heart of solid-state lighting and solid-state lasers. Thus, active semiconductor materials are a natural choice to realize efficient light-emitting all-dielectric nanoantennas and metasurfaces. 

A particularly attractive system allowing to integrate nanoscale emitters in Mie-resonant dielectric nanostructures are semiconductor quantum dots (QDs). QDs, also called artificial atoms, are semiconductor nanocrystals whose optoelectronic properties are dominated by the three-dimensional confinement of charge carriers. QDs are mainly famed for their exceptional light-emission properties. The emission wavelengths accessible with QDs cover an ultra-wide spectral band embracing the entire visible and near-infrared ranges. This is achieved not only by the choice of the constituent semiconductor material as in bulk semiconductors but also by tailoring the size of the QD and thereby the confinement energy. State-of-the-art QDs can have very high quantum efficiencies; they are stable and robust. Two types of QDs were so far considered as active elements of Mie-resonant dielectric nanostructures, namely colloidal QDs and self-assembled QDs. While the former can be realized in large quantities using wet chemistry approaches, the latter can be grown epitaxially with highest precision and can be directly integrated into monolithic active nanophotonic architectures. As such, both types of QDs offer their unique advantages for the use in light-emitting nanophotonics structures and devices. 

Various strategies were explored recently for integrating semiconductor QDs with Mie-resonant dielectric nanostructures. Note that such an integration naturally results in hierarchical nanostructures, combining feature sizes on the order of 1 nm for the electronic confinement (the QDs) with feature sizes on the order of 100 nm for the photonic confinement.

The simplest strategy, which was also the first to be demonstrated, is to cover as-fabricated dielectric nanostructures by a (usually polymeric) layer containing suspended QDs. Such layers can, e.g., be easily applied using spin-coating of the QD suspension. Using this method, spectral shaping of the emission spectrum of colloidal PbS QDs in a polystyrene matrix was demonstrated using arrays of Mie-resonant Si nanocylinders \cite{Staude2015} [see Fig.\,\ref{fig2}\,(a)]. The spectral density of emission showed a clear correlation with the spectral positions of the structure's Mie-type resonances, when the latter were systematically varied by tuning the diameter of the nanocylinders [see Figs.\,\ref{fig2}\,(b)].

However, in order to be able to carefully tailor the emission properties of the QDs coupled to the Mie resonators, a precise placement of the QDs with respect to the nanoresonators is highly desirable. One strategy to place QDs only in certain areas of the pre-fabricated nanoresonators employs selective chemical functionalization of the nanostructure surface \cite{Staude2014, Peter2017}. Electron-beam lithography is usually used to define small areas on or near the nanostructures, which are then exposed by wet chemical methods to a linker molecule. Subsequently these linker molecules selectively bind to suitably functionalized QDs. Based on this method, directional emission from a small cluster of colloidal QDs using a dielectric hafnium dioxide truncated-waveguide nanoantenna was demonstrated~\cite{Peter2017} [see Figs.\,\ref{fig2}\,(c-e)].

Very recently, the defined integration of colloidal QDs with dielectric nanoresonators was also demonstrated using dip-pen lithography~\cite{daewood2018}. In this method, the QDs are first dispersed in an ink, which is then deposited onto the nanoresonators using an atomic force microscopy (AFM) tip.

All methods for integrating QDs and dielectric nanoresonators discussed so far result in the QDs placed in the immediate nanoresonator environment. However, in order to access the strongest electromagnetic near-field enhancement it would be desirable to integrate the QDs directly {\it inside} the nanoresonators. Nanostructuring of epitaxially grown wafers incorporating self-assembled QDs offers this possibility. This was recently demonstrated for individual Mie resonators by Rutckaia {\it et al.}~\cite{Rutckaia2017}, where resonant Si nanopillars incorporating self-assembled Ge QDs were fabricated and optically characterized. Substantial enhancement of QD photoluminescence (PL) associated with a reshaping of the QD emission spectra was demonstrated. A similar strategy was used in \cite{yuan2017}, where Ge QDs were integrated in asymmetric silicon metasurface structures exhibiting Fano resonances with a high quality factor of 1011 [see Fig.\,\ref{fig2}\,(f-h)]. Over three orders of photoluminescence enhancement was observed. However, due to the indirect nature of the Ge electronic bandgap, the intrinsic quantum efficiency of such QDs is rather low. In order to further enhance the efficiency of light sources based on self-assembled QDs incorporated into Mie resonators, a material platform based on direct-bandgap compound semiconductors as commonly used in light emitting devices and semiconductor lasers is an obvious choice. Strong enhancement of emission from self-assembled InAs ODs integrated into Fano-resonant GaAs metasurfaces was recently demonstrated by Liu et al.~\cite{liu2017,liu2017b} Furthermore, using metasurfaces composed of Mie-resonant GaAs nanoresonators incorporating InAs QDs, spectral and directional shaping of light emission was achieved \cite{liu2017b,vaskin2018}. The case of a high-Q Fano-resonant metasurface consisting of asymmetric GaAs nanoresonators incorporating InAs QDs was also studied. For this case, large enhancements, spectral tailoring and lifetime shortening of the QD photoluminescence was observed \cite{liu2017}.
Apart from epitaxial structures, hierarchical metasurfaces consisting of SiO$_2$ nanocylinders with embedded luminescent Si nanocrystals were also studied along similar lines \cite{Capretti2017}.\\
While QDs offer distinct advantages over bulk semiconductors, most prominently the opportunity to tailor the emission wavelength and an enhanced efficiency by spatial confinement of the charge carriers inside the QDs, they add additional complexity to the system. Furthermore, they may suffer from inhomogeneous broadening where the emission wavelength of two distinct QDs is not exactly the same due to small size variations. As such, using bulk semiconductors as active components in Mie-resonant dielectric nanostructures also presents a viable option. An early work exploring this option investigated enhancement of bulk semiconductor bandgap photoluminescence in germanium waveguide cavity resonators on silicon at around 1.55 $\mu$m wavelength, showing about 30-fold enhancement in the collected spontaneous emission per unit volume when compared to a continuous germanium film of the same thickness \cite{Celebrano2015}. This was attributed to a combination of excitation enhancement at the pump wavelength, Purcell effect at the emission wavelength, and a beaming effect by the nanoresonators. More recently, lasing was demonstrated in arrays of gallium arsenide (GaAs) nanopillars supporting vertical electric dipole resonances \cite{Ha2018} [see Fig.\,\ref{fig2}\,(i-k)]. By designing one period of the array to support diffractive orders, a leaky channel is opened while preserving sufficiently high Q factors for lasing. The leaky resonance wavelength can be tuned via the geometrical parameters of the array such as particle size or lattice period, while the gain spectrum of the GaAs can be shifted by controlling the temperature. Combination of these two parameters enables lasing at different wavelengths and under different emission angles. Note that the un-passivated GaAs used as active material in this work provides only relatively low gain, and cooling to 200 K or less was required in order to observe lasing action.\\

\begin{figure*}
\centering
\includegraphics[width=0.9\linewidth] {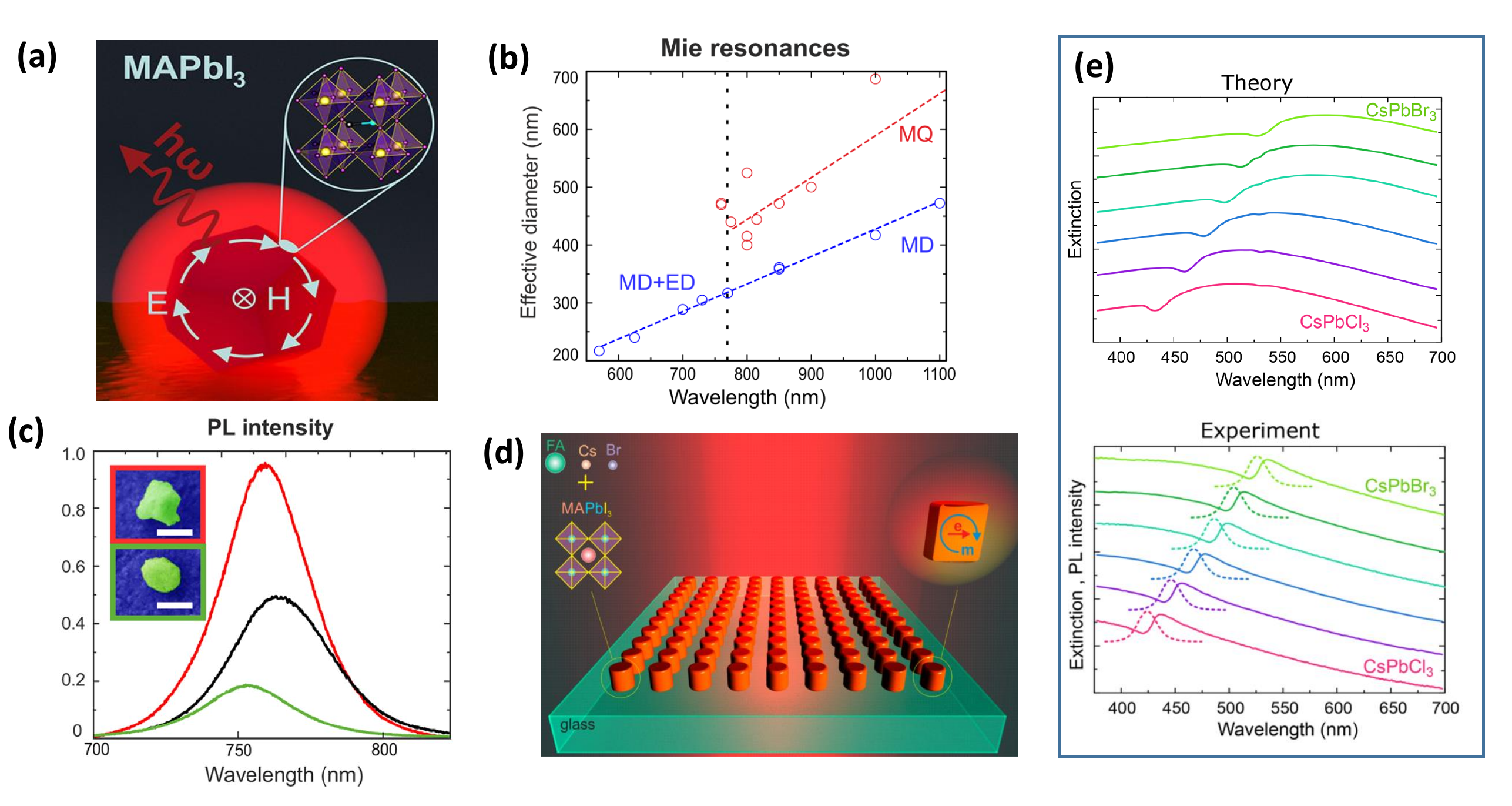}
\caption{{\bf Perovskite nanoparticles and metasurfaces.} (a) Schematic of tunable light-emitting halide perovskite nanoantenna with Mie-enhanced light emission depending on its chemical structure. (b) Experimental (open circles) and theoretical (dashed lines) spectral positions of the resonances vs. the diameter of perovskite nanoparticles~\cite{makarov2018}. 
(c) PL spectra normalized to the volume for two perovskite nanoparticles of different sizes, and compared with the PL spectrum of a 0.5~$\mu$m perovskite film. Scale bar in the SEM images is 400~nm~\cite{makarov2018}. (d) General concept of a light-emitting metasurface with enhanced functionalities achieved via structuring a perovskite film, optimized by alloying the organic cations for improved stability~\cite{makarov2017multifold}. (e) Theoretically calculated
and experimentally measured extinction spectra of nanoparticles with chemically tuned anion. Lorentzian-like curves around each spectrum show the corresponding measured PL spectrum for each doping concentration. The curves are shifted vertically for clarity.
}
\label{fig3}
\end{figure*}

\section{Halide perovskites}

Resonant nanostructures with low losses could be effective nanoscale light sources. However, the costs of multistage fabrication of nanostructures made of active inorganic semiconductors is a limiting factor for their practical implementation. Organic-inorganic (hybrid) perovskites of the MAPbX$_3$ family, where methylammonium (MA) stands for CH$_3$NH$_3$, and X stands for either I, Br or Cl, are a class of dielectric materials with excitonic states at room temperature, possessing high refractive indices, low losses at the exciton wavelength, and high quantum yield~\cite{sutter2015high}. These properties make them perfect candidates for effective nanoscale light sources. Superior material properties, along with low fabrication cost (such as wet-chemistry or spin coating depositions), allow for creating hybrid perovskite active metasurfaces~\cite{makarov2017multifold, gholipour2017organometallic} and microparticles~\cite{tang2017single}. Additionally, the emission wavelengths of hybrid perovskites can be gradually tuned over the entire visible range by simple replacing or mixing the anion compound (I, Br or Cl).

Tiguntseva {\it et al.}~\cite{makarov2018} demonstrated experimentally how to employ organic-inorganic halide perovskites, that support excitons at room temperature with high binding energies, to overcome limitations of semiconductor nanoantennas and create active nanoantennas with strong photoluminescence enhanced by coupling of excitons to dipolar and multipolar Mie resonances, as presented schematically in Fig.\,\ref{fig3}\,(a).  They employed simple technological processes based on laser ablation of a thin film prepared by a wet-chemistry method as a novel cost-effective platform for the fabrication of perovskite nanoparticles, active nanoantennas and nanostructures.

The study of white-light scattering from the perovskite nanoparticles with different diameters reveals their resonant behavior in both visible and infrared ranges. A nanoparticle with the diameter of 415~nm (measured by SEM) exhibits a pronounced scattering maximum around the spectral position of the exciton line ($\lambda$$\approx$770~nm). According to the multipole analysis, the experimentally obtained spectrum can be  described theoretically by several Mie resonances, namely MD, ED, MQ, and EQ, which contribute to the dark-field spectra in the low-loss range (i.e. $\lambda$$>$750~nm). 

Theoretical results for light scattering by a MAPbI$_3$ sphere reveal a variation of the Mie resonances in a broad range of nanoparticle diameters (200--700~nm). All resonances demonstrate a red shift with an increase of the nanoparticle diameter. By measuring the dark-field spectra from perovskite nanoparticles of different sizes, Tiguntseva {\it et al.}~\cite{makarov2018} found the same dependencies for distinguishable MD and MQ modes, as shown in Fig.\,\ref{fig3}(b). Figure~\ref{fig3}(c) shows the photoluminescence spectra normalized to the volume for perovskite nanoparticles of different sizes, as shown in the inset with the SEM images and marked by the corresponding colors of the frames.

Coupling of an exciton to a resonant nanostructure may significantly alter the optical response of the entire system. When a narrow exciton resonance of any material couples to a broader cavity resonance, one can expect an interference effect accompanied by the appearance of a characteristic Fano profile in the spectrum. Recently, the first experimental observation of a hybrid Fano resonance has been reported for both isolated perovskite nanoparticles and nanoparticle arrays~\cite{fano_pero} [see Fig.\,\ref{fig3}\,(e)]. Such Fano resonances originate from the coupling of perovskite excitons to the geometry-driven Mie resonances, and they are observed in the dark-field scattering and extinction spectra. The unique material properties of the perovskite nanoparticles allow to tune chemically their excitonic states over a range of 100~nm in a reversible way. These findings may have far-reaching implications for the design of perovskite-based optoelectronic devices as well as on-chip integrated reconfigurable light-emitting nanoantennas.

Perovskites can be employed for improving the characteristics of metasurfaces, especially for solar-cell applications, similar to the concept presented in Fig.\,\ref{fig3}\,(d), or as a novel type of metasurfaces based on nanoimprinted perovskite films optimized by alloying the organic cation part of perovskites. Makarov {\it et al.}~\cite{makarov2017multifold} revealed that such nanostructured perovskite films can exhibit a significant enhancement of both linear and nonlinear photoluminescence (up to 70 times) combined with advanced stability. These results suggest a cost-effective approach based on nano-imprint lithography combined with simple chemical reactions for creating new functional metasurfaces, which may pave the way towards highly efficient planar optoelectronic metadevices.

\section{Transition metal dichalcogenides}

Among the numerous new materials discovered by research in material science at the edge of nanotechnologies, atomically thin TMDC materials offer appealing optoelectronic properties. This gave rise to highflying expectations for applications in e.g. light detection, light emission, and light modulation. Already the sheer existence of such stable two-dimensional (2D) matter states in our three-dimensional world provides ample motivation for fundamental research. Moreover, the 2D semiconducting properties of these TMDCs with a bandgap in the visible to near-infrared range give rise to many expectations for applications. It appears that the optoelectronic properties of TMDCs are most attractive when they come as a single compound semiconductor layer since they possess a direct bandgap in this limit. The direct bandgap ensures momentum matching of photons and excitons without the need of additional phonon absorption, thus enabling strong photon-exciton interaction. Such TMDC monolayers are constructed from a chalcogenide sublattice formed by two parallel layers of hexagonally arranged S or Se atoms. Sandwiched between these two layers there is an additional hexagonal sublattice of transition metal atoms (either Mo or W). Since these two sublattices are displaced with respect to each other, the resulting broken inversion symmetry of the atom's arrangement gives rise to interesting effects. First, there arises a second-order nonlinear response of TMDCs, outclassing the $\chi ^{(2)}$ of bulk materials by orders of magnitude. Second, the combination of inversion symmetry breaking with strong spin-orbital coupling results in valley-selective circular dichroism, which can be exploited for applications in valleytronics. In addition to the interesting properties of TMDC monolayers, ensembles of such layers as well as ensembles of layers from different TMDC materials, which are just weakly bound by Van der Waals forces, offer further perspectives for the engineering of optoelectronic properties.

\begin{figure*}
\centering
\includegraphics[width=0.9\linewidth] {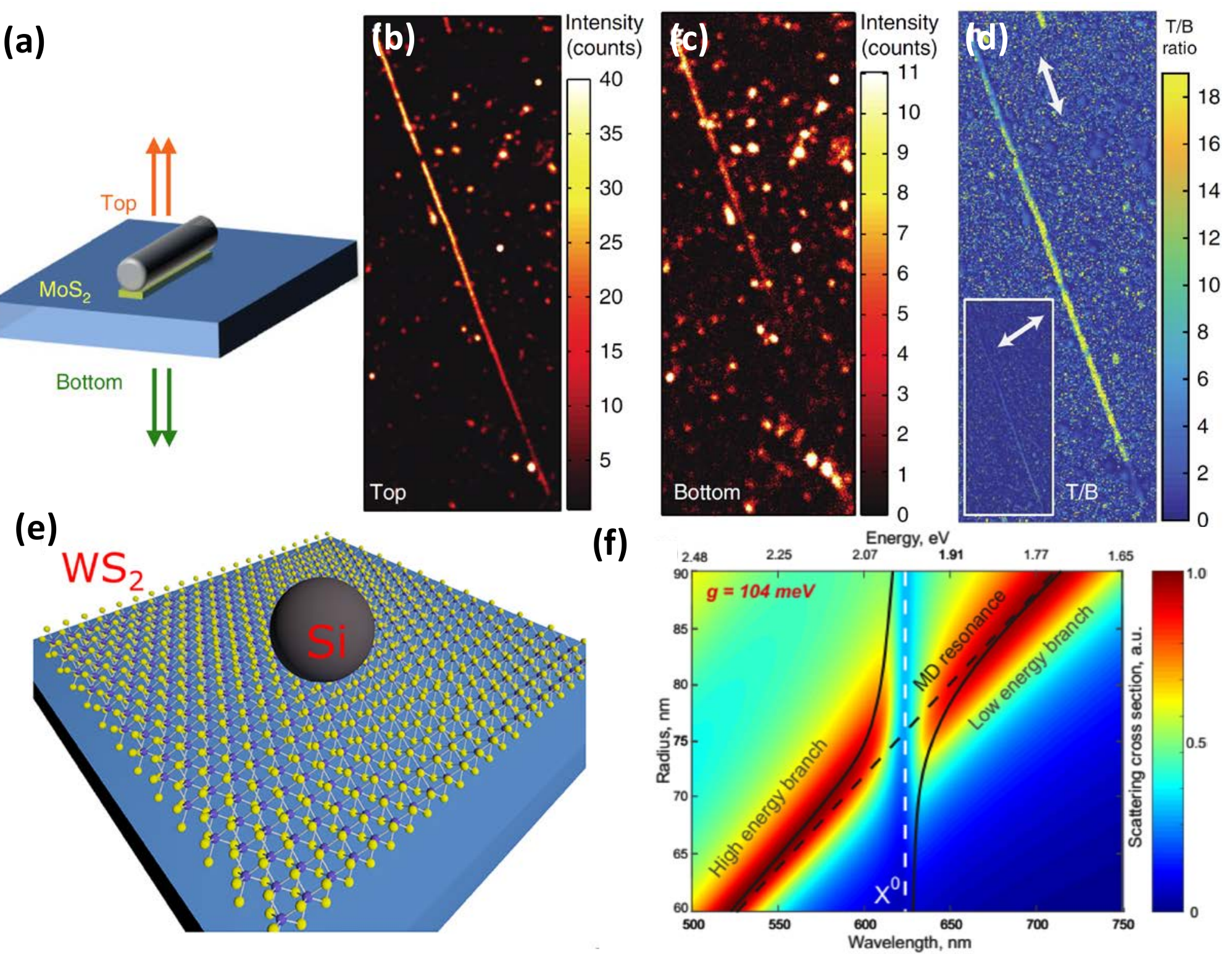}
\caption{{\bf Nanostructures with TMDC layers.} (a) Schematic of a Si nanowire placed on a MoS$_2$ monolayer patch, both attached to a sapphire substrate. (b-d) Experimental fluorescence images for transverse magnetic polarization for a Si nanowire having a length of 40 µm over which the diameter gradually changes from 20 to 40 µm. This gentle tapering of the nanowire allowed observing size dependent scattering directly from one sample. (b) shows scattering to the top, (c) to the bottom, and (c) the top to bottom (T/B) ratio, where the arrow indicates the electric field polarization direction for the collected light. The inset shows the T/B ratio for the other polarization direction demonstrating a clear enhancement of the T/B ratio connected to the excitation of optical resonances in the Si nanowire~\cite{Cihan_NP_2018}. (e) Schematic view of a sample comprising a single Si nanoparticle on a WS$_2$ monolayer. (f) Simulated scattering cross section of a Si nanoparticle covered entirely by a WS$_2$ monolayer immersed in water as a function on wavelength and the radius of the Si nanoparticle~\cite{LEPESHOV_ACSAMI_2018}.
}
\label{figTMDC}
\end{figure*}

Here we are concentrating our discussion on recent activities and perspectives to realize hybrid exciton-polariton systems formed by TMDC monolayers coupled to low loss dielectric nanoresonators. While published achievements in this direction are still rare owing to the immense complexity of the field, this is a particularly attractive approach for a multitude of reasons. First, due to the reduced dielectric screening and strong Coulomb interaction of electrons and holes the dipolar excitonic transition of TMDC monolayers with direct bandgap has a high oscillator strength. Despite of the small volume of the monolayers this gives rise to strong interaction with light, reaching easily 5-20\% for excitonic absorption, which, for comparison, outperforms a single semiconductor quantum well by a factor 10. Second, light-matter interactions with TMDCs can be further enhanced by the coupling to the photonic resonances of dielectric nanoresonators, since the resonant light concentration in the near-field can be engineered to overlap strongly with the small volume TMDCs. Engineering of this direct near-field coupling with the multiple electric and magnetic resonances and their interaction with the far-field offers many degrees of freedom to engineer the interaction of the entire hybrid system with the far-field. This concerns the selectivity with respect to direction, polarization, and spectrum. Third, the TMDC monolayers can be brought in direct contact with the dielectric nanoresonators without suffering from detrimental quenching of the dipole transitions. Similar to the already discussed quantum dots, the TMDC's bright excitons are prone to quenching, when being exposed to a high density of electronic states and the resulting dominance of the non-radiative decay channels. Previously this was the origin of many problems when coupling TMDCs to plasmonic nanoresonators, which can even become worse by unwanted doping from hot-electron injection from the plasmonic excitations. In contrast, the intrinsic optoelectronic properties of TMDCs can be broadly tuned by the contact to surrounding materials. Forth, the quite easy as well as inexpensive growth and transfer processes for TMDCs, which emerged in only a few years of research, makes them ideal candidates for integration with other optoelectronic materials and nanotechnologies for realizing complex and densely integrated systems. This list could be extended almost arbitrarily but should suffice for sketching the high potential of this new class of material when being hybridized with dielectric photonic nanoresonators. Thus, we expect a rapid growth of activities in this area in the near future. For further discussion of the properties of TMDCs we refer to ref. ~\cite{MAK_NP_2016}.

Recent demonstrations of coupling TMDCs to dielectric nanoresonators include the work by Cihan {\it et al.} on the control over the directionality, polarization state, and spectral emission relying on coherent coupling of the dipolar exciton emission from MoS$_2$ to the optical resonances of Si nanowires of different sizes~\cite{Cihan_NP_2018}. As sketched in Fig.\,\ref{figTMDC}\,(a), the system was prepared by a rather easy process starting from a CVD grown MoS$_2$ on sapphire substrate onto which tapered Si nanowires hand been drop casted. In the subsequent Ar plasma etching of the MoS$_2$ monolayer, the Si nanowires acted as a shadow mask. With this very simple nanoscale configuration, the authors could demonstrate a forward-to-backward ratio of 20 for the electric dipole emission to the far-field at a wavelength of 680 nm (see Fig.\,\ref{figTMDC}\,(b-d)). Such a property could be realized previously only by bulky optical systems. Moreover, they could demonstrate a wavelength tuning of the emission over 60 nm by changing the diameter of the nanowire. Besides these experimental achievements, the authors developed a versatile and intuitive modification of the Mie theory to describe directly the scattering of dipole emission by a nanoresonator instead of referring to plane wave scattering as in the classical Mie theory.

Another demonstration of a monolayer-based hybrid exciton-polariton system was reported by Lepeshov {\it et al.}, who investigated Si nanoparticles coupled to monolayers of WS$_2$~\cite{LEPESHOV_ACSAMI_2018}. As sketched in Fig.\,\ref{figTMDC}\,(e) they placed Si nanoparticles directly onto a flat monolayer of WS$_2$ and measured the photoluminescence spectra when immersing the configuration into different solvents. From their experimental findings the authors extrapolate to the prediction of a strong coupling regime, which could be achieved for a Si nanoparticle covered completely with a monolayer of WS$_2$ giving rise potentially to a Rabi splitting of more than 200 meV (see Fig.\,\ref{figTMDC}\,(f)). Such strong coupling could be achieved only by exploiting the symmetry matching of the magnetic dipole Mie mode of the Si nanoparticle to the exciton of the WS$_2$ and by a large dielectric constant of the surrounding medium. Their finding that the magnetic dipole mode of the nanoparticle couples much stronger to the bright exciton of the WS$_2$ is based on the dominant field component of the magnetic mode being tangential to the nanoparticle's surface and hence aligned with the exciton's dipole moment. In contrast, the electric dipole mode, even though it is much easier to excite from the far-field and it possesses a larger fraction of the mode's field outside the nanoparticle, does overlap with the TMDC's bright exciton dipole only weakly, due to the radial alignment of the mode. This finding will be important for future work, since the nice property that the dipole moments of bright excitations are aligned in the TMDC monolayers, as compared to the random orientation of colloidal quantum dots, can only be exploited when matched to the symmetry of the photonic mode. A similar investigation was recently reported by Bucher {\it et al.} who studied the emission characteristics from MoS$_2$ monolayers coupled to arrays of Si nanoantennas, which where structured by electron beam lithography on silicon thin-film wafers~\cite{BUCHER_CLEO_2018}.

\section{Other quantum emitters and further directions}

In addition to the active metadevices discussed above there is a rapidly growing number of works integrating also other quantum emitters into dielectric metadevices. In these approaches, scientists exploit the resonant character of dielectric nanostructures to enhance the emission of the quantum emitters and to control the emission characteristics with respect to space, angle, frequency, or polarization. The majority of these developments is still in a very preliminary state, where basic effects are in the center of interest, but it is already clear, that novel applications will arise from it.

\begin{figure*}
\centering
\includegraphics[width=0.9\linewidth] {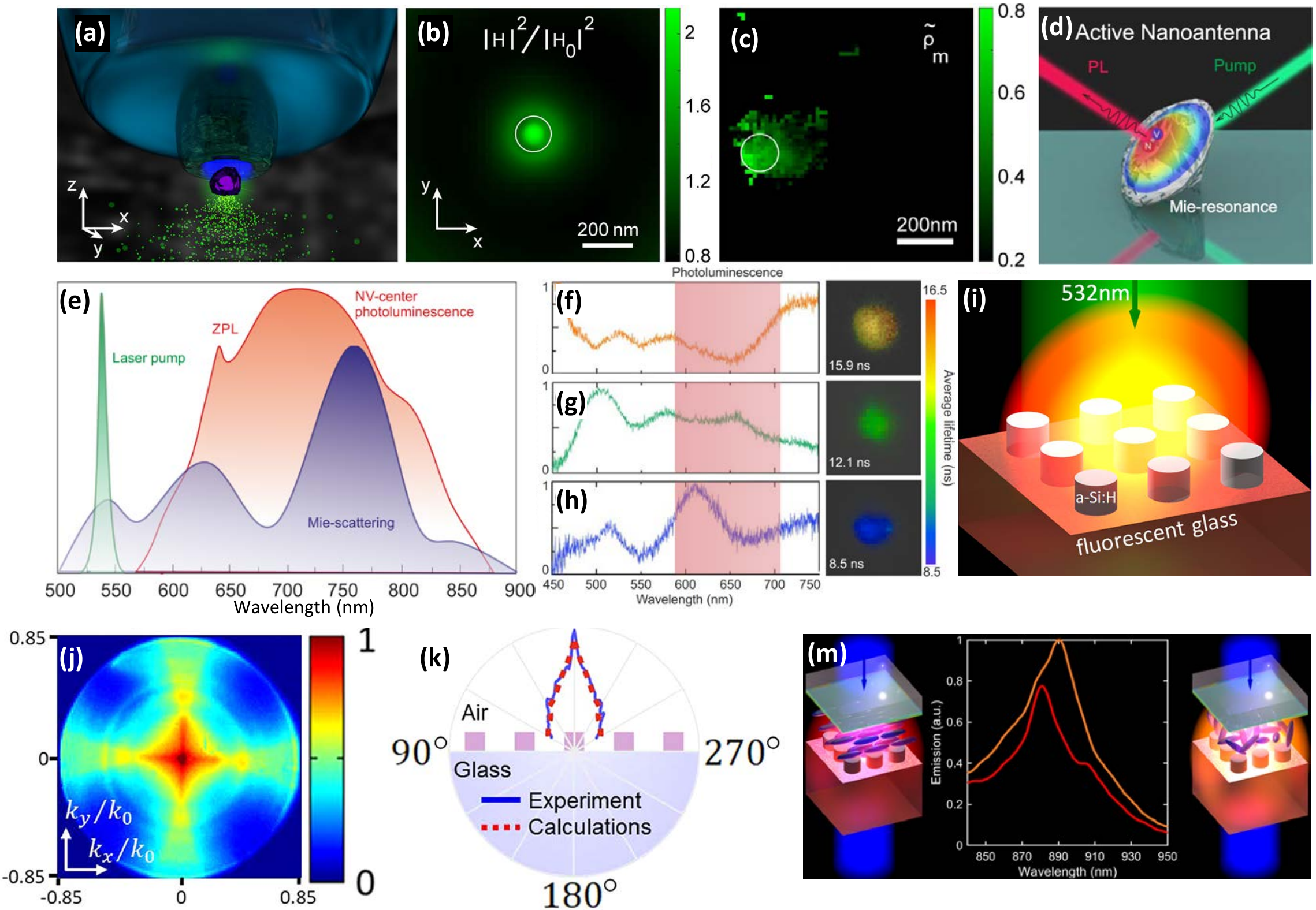}
\caption{ 
{\bf Other quantum emitters and further directions.} (a) Silicon nanocylinder antenna fabricated at the extremity of a near-field microscopy tip to control the antenna's relative position to a nanocrystal (purple particle) doped with trivalent europium ions possessing magnetic and electric dipolar transitions. The calculated field distribution of a magnetic mode supported by the nanocylinder at $\lambda$ = 590 nm is plotted in (a) and (b) in green. (c) Experimental mapping of the relative radiative magnetic local density of states surrounding the magnetic cylindrical antenna from (a). (a-c) taken from~\cite{sanzpaz2018}. (d) Scheme of a diamond nanoparticle acting as a multimode antenna. (e) Simulated spectral features of multiple Mie resonances of the diamond nanoparticles giving rise to resonant interaction both with the short wavelength pump light at 532 nm to excite the nitrogen vacancy (NV) centers in the nanoparticles as well as with the NV center's long wavelength photoluminescence (PL) and their characteristic zero phonon line (ZPL) at 637 nm. (f,g,h - left) Experimental dark-field scattering spectra of differently sized and shaped diamond nanoparticles. (f,g,h - right) Confocal luminescence maps with colors representing the lifetime of the photoluminescence, which is governed by the different spectral overlap of the scattering with the photoluminescence of the NV centers (shaded red area in spectra on the left). (d-h) taken from~\cite{zalogina2018}. (i) Scheme of Mie-resonant nanocylinders made from hydrogenated amorphous Silicon (a-Si:H) interacting with the intrinsic fluorescence of the glass substrate being excited by pump light illumination at 532 nm. (j) Experimental back focal plane emission image and (k) cross section of the emission pattern in the upper half-space of the Silicon nanocylinder arrays from (i), demonstrating the control of the substrate emission's directionality be the resonant interactions with the array. (i-k) taken from~\cite{vaskin2018}. (m) Scheme of a Silicon nanocylinder array integrated into a liquid crystal cell. By temperature induced switching of the liquid crystal from the nematic (left) to the isotropic (right) state, the emission spectrum (center) of the substrate's intrinsic fluorescence can be controlled dynamically (taken from~\cite{bohn2018}).
}
\label{figFURTHER}
\end{figure*}

First, we would like to underline the remarkable recent development of hybridizing all-dielectric resonant nanostructures with lanthanide ions. These constituents match ideally, since both support excitations beyond the electric dipole limit. Even though light-matter interaction is usually dominated by the electric field and the magnetic field's interaction is typically weaker by roughly 4 orders of magnitude, trivalent lanthanide ions like Eu$^{3+}$ or Er$^{3+}$ exhibit specific magnetic-dipole-dominated transitions. In addition, nanoscale dielectric resonators can strongly modify the magnetic Purcell factor by their magnetic dipole and higher-order modes~\cite{baranov2017}. Hence, the interaction of light with lanthanide ions close to dielectric resonators can be enhanced and engineered. This concept was recently explored experimentally by Sanz-Paz et al. who studied the emission from colloidal particles of YVO$_4$ doped with 20 percent Eu$^{3+}$ when placed in the vicinity of a cylindrical Si nanoantenna supporting a magnetic dipole resonance at the frequency of the magnetic dipole transition of Eu$^{3+}$~\cite{sanzpaz2018}. In their study, this magnetic emission was compared to the emission from the electric dipole transition of Eu$^{3+}$ when resonantly coupled to an electric dipole antenna constructed from an aluminum nanowire. The precise control of their relative position by mounting both nanoresonators on two tips of a near-field optical microscope allowed a systematic position dependent mapping of the magnetic vs electric interaction (see Fig. 5(a-f)). Even though this study was conducted for single antennas and single doped particles, it clearly paves the way for a broad exploitation of this effect in metasurfaces, where an entire array of dielectric nanoresonators is hybridized by coupling to an environment of lanthanide ions. A study implementing this concept was reported very recently by Noginova et al. ~\cite{noginova2018}. They have coupled arrays of Mie-resonant Si nanocylinders to Eu$^{3+}$ ions by covering the nanocylinders with a polymer layer which was doped with an Eu$^{3+}$ compound. Their systematic study included a change of the radius of the nanocylinders, by which the Mie resonances could be tuned. This allowed to clearly correlate the emission spectrum to the interaction of the electric and magnetic dipole transitions of the Eu$^{3+}$ with the individual Mie resonances. These two recent works provide just a glimpse of the rich world of new hybrid metadevices exploiting light matter interactions beyond the classical electric dipole limit, which up to now had been forming the basis of almost all active photonic systems.

The second stream of very initial studies, which we would like to highlight, concerns active metadevices exploiting the emission from the nitrogen-vacancy centers in diamond (NV centers). This is a particularly promising approach since the large epsilon and broad transparency spectrum (from VIS to IR) of diamond combined with the richness of light-matter interaction of diamond defect centers offer to integrate high-performance metadevices into a single material platform. This is in contrast to the examples of metadevices discussed so far, where the active component was added to the dielectric nanoresonators mainly by hybridization with another material. The large variety of emitting color centers, which can be created in diamonds, eventually allows addressing applications over the entire VIS to IR range. Among those, the negatively charged NV centers currently appear to hold the highest application promise due to their zero-phonon line emission at 637 nm, which can be observed in their photoluminescence even at room temperature. The phonon sidebands of this sharp line can be further suppressed by engineering of the photonic modes in the volume of the NV center. In the past nanodiamonds had been integrated with plasmonic nanoantennas. However, despite the strong emission enhancement resulting from the huge field concentration of plasmonic antennas, the metal-associated losses form a severe obstacle for any process exploiting the quantum nature of the absorbed and generated photons. Therefore, replacing the metal component in an all-dielectric meta-device offers an enormous potential for quantum application. Besides a vivid reflection of the approach to integrate NV centers into resonant nanodiamonds during the discussions at recent conferences, up to now we are aware only of a single published study by Zalogina et al., who reported about light emission from resonant diamond nanoparticles~\cite{zalogina2018}. As illustrated in Fig. 5(g) their diamond nanoparticles supported Mie resonances overlapping spectrally and spatially with the transition of the NV centers. In their detailed experimental study, they fabricated the diamond nanoparticles by milling diamond crystals, which they had obtained from chemical vapor deposition. Using Raman spectroscopy, they selected the nanoparticles with crystalline phase for further investigation, having different shapes and sizes. For dimensions ranging from several hundred nanometers to few microns, they could investigate nanoparticles supporting different higher order modes and study their interaction with the intrinsic defects, resulting in modified emission spectra. Approximating the different shapes by spheres, they could obtain good correspondence to a Mie model and hence, attribute the observed size dependence of the emission spectrum to the coupling of higher order Mie modes with the transitions of the NV centers, which are randomly distributed inside the nanoparticles. For resonant coupling, they could demonstrate a lifetime reduction by almost 3 times as compared to NV centers in nanodiamonds, which do not support photonic resonances. From this study one can conclude, that for a well-designed size and shape of diamond nanoparticles the excitation and luminescence rates of NV centers inside nanoparticles can be enhanced. In perspective, this would provide an ideal platform for metadevices serving a variety of applications, e.g. in bioimaging, sensing, and quantum signal processing. However exploiting the full potential of this technology platform will require the precise control of size and shape of the diamond nanoparticles as well as control over the position of the NV centers inside the nanoparticles.

Finally yet importantly, we would like to direct attention to already demonstrated application perspectives of active dielectric metasurfaces for beam shaping and even dynamic control of their emission properties. As a result of a recent study of the emission from a fluorescent glass substrate, which was covered by an array of Si nanocylinders, Vaskin et al. reported spectral and spatial shaping of the emission by the collective resonances of the nanoresonators forming a metasurface~\cite{vaskin2018b}. As sketched in Fig. 5(f), they coupled the intrinsic fluorescence from a glass substrate in the spectral range from 600 nm to 900 nm to a square array of Si nanocylinders at the surface of the glass substrate. The important novelty of their approach is the simultaneous exploitation of emission enhancement by the Mie-type magnetic dipolar resonances of the individual Si nanocylinders with the coherent scattering by the periodic arrangement of the nanocylinders on the metasurface. This gave rise to an emission mainly in a single direction out of the substrate plane. While still being a fundamental study relying for simplicity of the technological realization on the weak intrinsic fluorescence of glass, it clearly demonstrates the application potential of active metasurfaces for the creation flat sources of tailored light fields. Besides the demonstrated versatility to control the emission from active metadevices by the nanostructuring of dielectric surfaces, real applications will require dynamic control of the emission characteristics, as e.g. for displays. However, all previously discussed studies relied on static arrangements of nanoresonators or a basic mechanical control of positioning for single resonators in fundamental studies. An important next step for active metadevices was recently reported by Bohn et al. who achieved active tuning of the emission characteristics by combining active dielectric metasurfaces with the liquid crystal (LC) technology~\cite{bohn2018}. Similar to reference~\cite{vaskin2018b}, their demonstration is based on the intrinsic fluorescence from a glass substrate coupled to an array of Si nanocylinders. By exploiting the specific wavelength shift of the magnetic and electric dipolar Mie-type resonances of these Si nanocylinders by a change of the LC from the nematic state to the isotropic state, they modified the local photonic density of states (see Fig. 5(h)). For this basic demonstration, the change of state of the LC was realized by a simple temperature tuning of the entire device. Consequently, they could modify dynamically the spectral composition of the emission from the substrate. As a proof of the applicability for sheer amplitude modulation they demonstrated control of the emission amplitude at the central emission wavelength of $\lambda = 900 nm$ by a factor of 2.

\section{Summary and outlook}

We have presented a brief overview of the most recent advances in the emerging field of active dielectric nanophotonics driven by Mie-type resonances. This is a rapidly developing research direction with a great potential for applications in new types of light sources, next-generation displays, quantum signal processing and lasing. A subwavelength confinement of the local electromagnetic fields in resonant high-index dielectric photonic nanostructures due to Mie and Fano resonances, as well as the interference physics of the bound states in the continuum can boost many optical effects, thus offering novel opportunities for the subwavelength control of light in active nanostructures. This latest activity in all-dielectric resonant nanophotonics and meta-optics involves naturally active structures as components of future metadevices, which are defined as devices having unique and useful functionalities that are realized by structuring of functional matter on the subwavelength scale. 

One of the prospective future developments of the field can be associated with topological photonics~\cite{top1} and lasing of topologically robust systems~\cite{las1,las2}. Recent years have witnessed intense efforts toward exploiting these phenomena in dielectric photonic structures~\cite{top2}, and active meta-optics is expected to play a crucial role in the realization of these ideas in practical devices. It is expected that nonmagnetic topological insulator laser systems will exhibit topologically protected single-mode lasing being robust against defects, and operating at the nanoscale being driven by active topological platforms, enforcing predetermined unidirectional lasing without magnetic fields. 

Another future direction of light-emitting metasurfaces will be to enhance the complexity of the spatial emission patterns. In contrast to passive wavefront-shaping metasurfaces illuminated by a plane wave, however, one can usually not rely on coherence of the emitted light over the entire metasurface plane to generate spatially complex light fields. This is due to the fact that each emitter will emit largely independently when considering spontaneous emission. Spatially complex light fields may still be accessible by constructing metasurfaces from individually directional dielectric nanoantennas or by exploiting coupling between the meta-atoms. A natural future step will also be to combine active control of spontaneous emission \cite{bohn2018} with shaping of the spatial emission pattern \cite{vaskin2018,Cihan_NP_2018} to implement dynamic control of the latter. Also, the possibility to control the polarization of the light emitted by active Mie-resonant nanostructures remains underutilized.
Finally, further exploring different implementations of active Mie-resonant nanostructures will remain a key challenge, which may eventually open up electrical driving schemes, thereby rendering dielectric nanoantennas and metasurfaces practical for real-world applications in the next generation classical and quantum light sources, lighting technology and light-field displays. 

Moreover, meta-optics has an even broader meaning, highlighting the importance of optically induced magnetic response for many applications, including optical sensing, parametric amplification, fast spatial modulation of light, nonlinear active media, as well as both integrated classical and quantum circuitry and topological photonics, underpinning a new generation of highly-efficient and ultrafast metadevices.

\section*{Acknowledgments}

The authors acknowledge useful collaboration with many colleagues and students. IS and TP acknowledge support by the German Federal Ministry of Education and Research (project identifier 13N14147). TP was supported by the German Research Foundation (grant PE 1524/7-1) and the European Union (grant 2013-5659/002-001). IS gratefully acknowledges the financial support by the Thuringian State Government within its ProExcellence initiative (ACP$^{2020}$) and
by the German Research Foundation (STA 1426/2-1). YSK was supported by the Humboldt Foundation and the Strategic Fund of the Australian National University. 


\end{document}